\documentclass[a4paper,11pt]{article}
\usepackage{jheppub} 
\usepackage{lineno}
\usepackage{float}
\usepackage{slashed}
\usepackage{xfrac,bigints}
\usepackage{physics}
\usepackage{subfigure}

\title{Real-time chiral dynamics at finite temperature from quantum simulation}

\author[a,b]{Kazuki Ikeda,}
\emailAdd{kazuki.ikeda@stonybrook.edu}
\affiliation[a]{Co-design Center for Quantum Advantage, Stony Brook University, Stony Brook, New York 11794-3800, USA}
\affiliation[b]{Center for Frontiers in Nuclear Science, Stony Brook University,
Stony Brook, NY 11794, USA}

\author[c,d,b,e]{Zhong-Bo Kang,}
\emailAdd{zkang@physics.ucla.edu}
\affiliation[c]{Department of Physics and Astronomy, University of California,
Los Angeles, CA 90095, USA}
\affiliation[d]{Mani L. Bhaumik Institute for Theoretical Physics, University of California,
Los Angeles, CA 90095, USA}
\affiliation[e]{Center for Quantum Science and Engineering, University of California, Los Angeles, CA 90095, USA}

\author[a,b,f]{Dmitri E. Kharzeev,}
\emailAdd{dmitri.kharzeev@stonybrook.edu}
\affiliation[f]{Department of Physics, Brookhaven National Laboratory, Upton, New York 11973-5000, USA}

\author[g]{Wenyang Qian,}
\emailAdd{qian.wenyang@usc.es}
\affiliation[g]{Instituto Galego de F\'isica de Altas Enerx\'ias IGFAE, Universidade de Santiago de Compostela,
E-15782 Galicia-Spain}

\author[h]{and Fanyi Zhao}
\emailAdd{fanyi@mit.edu}
\affiliation[h]{Center for Theoretical Physics, Massachusetts Institute of Technology, Cambridge, MA 02139, USA}

\abstract{In this study, we explore the real-time dynamics of the chiral magnetic effect (CME) at a finite temperature in the (1+1)-dimensional QED, the massive Schwinger model. By introducing a chiral chemical potential $\mu_5$ through a quench process, we drive the system out of equilibrium and analyze the induced vector currents and their evolution over time. The Hamiltonian is modified to include the time-dependent chiral chemical potential, thus allowing the investigation of the CME within a quantum computing framework. We employ the quantum imaginary time evolution (QITE) algorithm to study the thermal states, and utilize the Suzuki-Trotter decomposition for the real-time evolution. This study provides insights into the quantum simulation capabilities for modeling the CME and offers a pathway for studying chiral dynamics in low-dimensional quantum field theories.}

\begin{document}
\maketitle
\flushbottom

\section{Introduction}
In recent years, the exploration of quantum simulations for field theories~\cite{Jordan:2011ci, Jordan:2012xnu, Jordan:2014tma} has garnered significant attention across many subfields of physics~\cite{Bauer:2022hpo, Bauer:2023qgm}. This growing interest is largely driven by the advancements in quantum computing technology, which promises to revolutionize our ability to tackle complex and computationally intensive problems. Traditional classical computers, despite their power, often fall short when dealing with the intricate dynamics and properties of quantum systems due to the exponential growth of computational resources required. On the other hand, quantum simulations hold the potential for breakthrough discoveries in areas such as high-energy physics, material science, and chemistry, aiding in understanding phase transitions~\cite{Czajka2022, Czajka:2022plx, Davoudi:2022uzo, Qian:2024xnr}, solving quantum many-body problems~\cite{Kreshchuk:2020dla,Qian:2021jxp, Barata:2023clv, Du:2023bpw}, and discovering new materials with unique properties~\cite{McArdle:2018tza}.

To explore these potential developments, we focus on the Schwinger model~\cite{Schwinger:1962tp}, quantum electrodynamics (QED) in (1+1) dimensions. 
The Schwinger model provides an excellent testing ground for studying various phenomena, including confinement, dynamical mass generation, and anomaly-induced symmetry breaking. It offers the simplicity of fewer degrees of freedom while still encapsulating the rich physics of gauge theories, making it an ideal benchmark model for quantum simulations on both quantum and classical hardware~\cite{Klco:2018kyo,Farrell:2023fgd,Farrell:2024fit,Butt:2019uul,Magnifico:2019kyj,Shaw:2020udc,Kharzeev:2020kgc,Ikeda:2020agk,2023arXiv230111991F,PhysRevD.107.L071502,Ikeda:2023zil,Ikeda:2023vfk, Florio:2024aix}.

In this study, we explore the real-time dynamics of the chiral magnetic effect (CME) at finite temperature within the massive Schwinger model. The CME refers to the generation of an electric current along a magnetic field in systems where there is an imbalance between left-handed and right-handed fermions~\cite{Kharzeev:2004ey, 2008NuPhA.803..227K, Fukushima:2008xe,Kharzeev:2013ffa,Kharzeev:2015znc,Landsteiner:2016led,Kharzeev:2020jxw,2016NatPh..12..550L}. This imbalance is typically produced by the presence of a chiral chemical potential. To investigate the CME within the context of the Schwinger model, we employ a quench protocol as outlined in our previous study~\cite{Kharzeev:2020kgc}. By introducing a chiral chemical potential through this quench process, we drive the system out of equilibrium and analyze the resulting vector currents and how they evolve over time. The Hamiltonian is adjusted to include the time-dependent chiral chemical potential, enabling the investigation of the CME within a quantum computing framework.

The interest in the study of the temperature dependence of the real-time CME stems from the following consideration. The CME is intrinsically a non-equilibrium, real-time phenomenon. However, when the external fields and the chiral chemical potential vary slowly with time, the CME response is driven by the fermion zero modes, and should thus be temperature-independent \cite{Fukushima:2008xe,PhysRevD.80.034028}. This independence of CME current on the temperature has been established in analytical computations \cite{Fukushima:2008xe,PhysRevD.80.034028}. However, under a rapid quench, the CME current receives contributions from excited fermion modes \cite{PhysRevD.80.034028}, and can thus develop a dependence on the temperature. This temperature dependence  has not been evaluated before from first principles. In this paper, we will evaluate it using the methods of quantum simulation in massive Schwinger model at finite temperature.  

We employ the quantum imaginary time evolution (QITE) algorithm to study the thermal states and utilize the Suzuki-Trotter decomposition for real-time evolution. The QITE algorithm is essential for efficient quantum simulation on quantum circuits for several reasons, and various related algorithms have been developed~\cite{PhysRevLett.98.070201,PhysRevB.78.155117,PhysRevB.91.115137}.
It enables the simulation of thermal states, crucial for studying phenomena like phase transitions at finite temperatures~\cite{Motta2020,Davoudi:2022uzo,Gomes:2021ckn,PhysRevLett.111.010401,McArdle2019,PhysRevC.109.044322,Yuan2019theoryofvariational}. Since imaginary time evolution involves non-unitary operations, which cannot be directly implemented on quantum hardware, QITE approximates these operations with unitary ones, making it feasible on quantum circuits. Additionally, traditional classical algorithms struggle to prepare thermal states of many-body systems with a
sign problem, while QITE allows for efficient exploration of these state spaces. Its versatility makes it applicable to a wide range of quantum systems and models, bridging theoretical models and practical implementations on quantum hardware. By iteratively applying small imaginary time steps and recalibrating the system, the QITE ensures the system gradually reaches the desired state where subsequent measurements are performed to extract physical observables.

Our motivation follows the successful quantum simulations performed on the (1+1)-dimensional Nambu–Jona-Lasinio (NJL) model~\cite{Czajka2022,Czajka:2022plx}, where quantum algorithms provided insights into the chiral phase transition at finite temperature and chemical potential. Extending this line of research, our previous work on the quantum computational study of field theories has demonstrated the feasibility and efficacy of these simulations in delivering significant theoretical insights. This work on the Schwinger model bridges the study of simple models accessible by current quantum hardware and the more complex theories underpinning the standard model of particle physics. By exploring the unique properties and dynamics of the Schwinger model, we aim to provide insights into chiral dynamics in low-dimensional quantum field theories.

The rest of this article is organized as follows. In Sec.~\ref{sec:model}, we present the model and the quench protocol that we study in this work. In Sec.~\ref{sec:QITE}, we describe our algorithm of the quantum imaginary time evolution to implement the thermal states and subsequent expectation evaluation to extract physical observables at finite temperature. In Sec.~\ref{sec:results}, we show the results from our quantum simulations, where we discuss the real-time chiral dynamics of the current at finite temperatures, the phase diagram of the Schwinger model, and the temperature dependence of the electric charge (axial current). Finally, the article is concluded in Sec.~\ref{sec:conclusion}.

\section{\label{sec:model}The massive Schwinger model with the topological term} 
\subsection{The Schwinger Hamiltonian}
The action of the massive Schwinger model~\cite{Schwinger:1962tp} with $\theta$ term in $(1+1)$-dimensional Minkowski space is
\begin{align}
 S = \int d^2x\left[-\frac{1}{4} F^{\mu\nu} F_{\mu\nu} + \frac{g\theta}{4\pi}\epsilon^{\mu\nu}F_{\mu\nu} + \bar{\psi}(i\slashed{D}-m)\psi\right],
\end{align}
with $\slashed{D}=\gamma^\mu(\partial_\mu + i gA_\mu)$.
Here, $A_\mu$ is the $U(1)$ gauge potential, $E=\dot{A}_1$ is the corresponding electric field, $\psi$ is a two-component fermion field, $m$ is the fermion mass, and $\gamma^\mu$ are two-dimensional $\gamma$-matrices satisfying the Clifford algebra. When $g\neq0$, the $\theta$-term induces the classical background electric field, which breaks the parity symmetry. The gauge field $A_\mu$ and the coupling constant $g$ have mass dimensions 0 and 1, respectively.
By a chiral transformation, $\psi\to e^{i\gamma^5\theta}\psi$ and $\bar{\psi}\to \bar{\psi}e^{i\gamma^5\theta}$, the action is transformed to 
\begin{align}
\label{action}
 S = \int d^2x\left[-\frac{1}{4} F^{\mu\nu} F_{\mu\nu}  + \bar{\psi}(i\gamma^\mu D_\mu-me^{i\gamma^5\theta})\psi\right].
\end{align}
The action is invariant under this transformation only up to the boundary term.
It is evident from eq.~\eqref{action} that the massive theory with a positive mass $(m>0)$ at $\theta=\pi$ is equivalent to the theory at $\theta=0$ but with a negative mass $(-m)$.  

The Hamiltonian in the temporal gauge $A_0=0$ is given by
\begin{align} \label{eq:Ham_conti}
H=\int dz \Big[    
    \frac{E^2}{2}
    -\bar{\psi}(i\gamma^1\partial_1 - g\gamma^1A_1 - me^{i\gamma^5\theta})\psi \Big] \,,
\end{align}
where the space-time coordinate is labeled by $x^\mu=(t,z)$. We denote the Pauli matrices as $X$, $Y$, and $Z$, and use the following convention for the Dirac matrices: $\gamma^0 = Z$, $\gamma^1 = i\,Y$, $\gamma^5=\gamma^0 \gamma^1 = X$.
In $(1+1)$ dimensions, the axial charge density $Q_5(x)\equiv\bar{\psi}(x)\gamma^5\gamma^0\psi(x)$ 
and the vector current density $J(x)\equiv\bar{\psi}(x)\gamma^1\psi(x)$ are related 
by $Q_5(x) = -J(x)$. Moreover the vector charge density $Q(x)\equiv\bar{\psi}(x)\gamma^0\psi(x)$ and the 
axial current density $J_5(x)\equiv\bar{\psi}(x)\gamma^5\gamma^1\psi(x)$ are related by $Q(x) = J_5(x)$.

Typically, the CME refers to the generation of electric current induced by the chiral imbalance in a background magnetic field. Though magnetic field is absent in one spatial dimension, it is known that in the limit of a strong magnetic field a dimensional reduction to the (1+1) dynamics occurs~\cite{PhysRevD.83.085007}. In this effective (1+1) dimensional theory, the CME is indeed the generation of electric current by the chiral imbalance that we discuss in the charge densities. The dimensional reduction requires that the gap to the first excited Landau level is much larger than the temperature, so the domain of applicability of our effective (1+1) dimensional theory is $eB\gg T^2$, where $B$ is magnetic field and $T$ is the temperature.

\subsection{\label{sec:quench}Real-time chiral dynamics at finite temperature}

As we have mentioned in the introduction, the chiral magnetic effect refers to the generation of an electric current along a magnetic field in a system with an imbalance in the number of left-handed and right-handed fermions, induced by the presence of a chiral chemical potential. In the Schwinger model, when the chiral quench is applied, the system is driven out of equilibrium, creating a non-zero chiral chemical potential $\mu_5$. This chiral chemical potential represents an asymmetry between left-handed and right-handed fermions.

Here, we consider the $\mu_5$-quench~\cite{Kharzeev:2020kgc} protocol: The system Hamiltonian with $\theta=0$ is prepared at time $t<0$. Then, starting at $t=0$, the Hamiltonian rotates the $\theta$ angle according to $\theta=-2\mu_5t$ corresponding to constant chiral chemical potential (i.e. $\dot{\theta} = -2\mu_5$.). The Hamiltonian after the quench is modified as
\begin{align} \label{eq:Ham2}
H'=\bigintsss dz \Bigg[    
    \frac{E^2}{2}
    -\bar{\psi}\left(i\gamma^1\partial_1 - g\gamma^1A_1-\gamma^1\frac{\dot{\theta}}{2} - me^{i\gamma^5\theta}\right)\psi \Bigg] \,.
\end{align}
The real-time evolution of an operator $O$ can be evaluated by the time-ordered integral 
\begin{equation}
\label{eq:real_time_evolution}
    O(t)=\mathcal{T}[e^{i\int_0^tdt'H'(t')}]O\mathcal{T}[e^{-i\int_0^tdt'H'(t')}],
\end{equation}
and the thermal expectation value of $O$ evaluated at time $t$ at finite temperature $T$ is
\begin{equation}\label{eq:thermal_observable}
    \langle O(t)\rangle_\beta=\frac{\Tr[e^{-\beta H}O(t)]}{\Tr [e^{-\beta H}]},
\end{equation}
where $H$ is the unquenched Hamiltonian in eq.~\eqref{eq:Ham_conti} at $\theta=0$ and $\beta=1/T$ is the inverse of the temperature of the system.

Notably, this evolving $\theta = -2\mu_5 t$ after the quench at $t = 0$ indicates a dynamical chiral chemical potential, driving the system into a non-equilibrium state where left-handed and right-handed fermions are imbalanced. In the presence of this chiral chemical potential, a vector current $J$ (which is typically zero before the quench) is induced. The vector current's time evolution at a finite $\beta$, influenced by the quench, aligns with the characteristic behavior observed in the CME at a finite temperature.

\subsection{The lattice Hamiltonian on the qubits}
To discretize our Hamiltonian, we use staggered fermions~\cite{Kogut:1974ag, Susskind:1976jm} on a lattice such that
\begin{align}
    \psi(x\equiv na) = \frac{1}{\sqrt{a}} \left(
    \begin{array}{cc}
    &\chi_{2n}
    \\
    &\chi_{2n+1}
    \end{array}
    \right)\,,
\end{align}
where $a$ is the finite lattice spacing. Then the lattice Hamiltonian corresponding to eq.~\eqref{eq:Ham2} is~\cite{Ikeda:2020agk} 
\begin{align}
\begin{aligned}
\label{eq:lattice_total_ham3}
H &= \frac{a g^2}{2}\sum_{n=1}^{N-1} L_n^2
 -\frac{i}{2a}\sum_{n=1}^{N-1}
 \big[\chi^\dag_{n+1} U_n\chi_{n}-\chi^\dag_{n}U_n^\dag\chi_{n+1}\big]
+ m\cos\theta\sum_{n=1}^{N} (-1)^n \chi^\dag_n\chi_n\\
 &+ i\frac{m\sin\theta}{2}\sum_{n=1}^{N-1} (-1)^n \big[\chi^\dag_{n+1} U_n\chi_{n}-\chi^\dag_{n}U_n^\dag\chi_{n+1}\big]-\frac{\dot{\theta}}{4a}\sum_{n=1}^{N-1}\big[\chi^\dag_{n+1}\chi_{n}+\chi^\dag_{n}\chi_{n+1}\big]\,,
\end{aligned}
\end{align}
where $L_n$ is the electric field operator satisfying the Gauss' law constraint
\begin{equation}
    L_{n}-L_{n-1} =  \chi_n^\dagger \chi_n-\frac{1-(-1)^n}{2}.
    \label{eq:gauss_staggered}
\end{equation}
Using the Gauss' law constraint, we eliminate the link fields $U_n$ by the gauge transformation,
\begin{align}
    \chi_n\to g_n\chi_n,\quad
    \chi^\dag_n\to \chi^\dag_n g_n^\dag,\quad
    U_n \to g_{n+1}U_n g_n^\dag,
\end{align}
with
\begin{align}
    g_1=1, \quad g_n = \prod_{i=1}^{n-1}U^\dag_{i}.
\end{align}

For the purpose of quantum simulation, we rewrite the lattice Hamiltonian in the spin representation using the Jordan--Wigner transformation~\cite{Jordan:1928wi}:
\begin{align}
\begin{aligned}
 \chi_n=\frac{X_n-iY_n}{2}\prod_{i=1}^{n-1}(-i Z_i).
\end{aligned}
\end{align}
The full Hamiltonian in the qubit representation considered in this study is
\begin{align}
\begin{aligned}
\label{eq:Ham}
H=&\sum_{n=1}^{N-1}\left(\frac{1}{4a}-\frac{m}{4}(-1)^n\sin\theta\right)\Big[X_n X_{n+1}+Y_n Y_{n+1}\Big]
     +\frac{m\cos\theta}{2}\sum_{n=1}^N(-1)^n Z_n\\
     &+\frac{a g^2}{2}\sum_{n=1}^{N-1}L^2_n-\frac{\dot{\theta}}{8a}\sum_{n=1}^{N-1}\Big[X_n Y_{n+1}-Y_n X_{n+1}\Big].
 \end{aligned}
 \end{align}
Moreover, the local vector and axial charge densities are, respectively~\cite{Ikeda:2023vfk}, 
\begin{align}
    Q_n \equiv \,& \bar{\psi}\gamma^0\psi = \frac{Z_n+(-1)^n}{2a},\\
    Q_{5,n} \equiv \,& \bar{\psi}\gamma^5\gamma^0\psi = \frac{X_nY_{n+1}-Y_nX_{n+1}}{4a}\,.
\end{align}
For later convenience, we introduce the average vector and axial charge operators that sum over all lattice sites,
\begin{align}
    Q\equiv\frac{a}{N}\sum_{n=1}^{N}Q_n, \qquad
    Q_5\equiv\frac{a}{N-1}\sum_{n=1}^{N-1}Q_{5,n}.
\end{align}
Notably, $Q$ commutes with the Hamiltonian while $Q_5$ does not. With the boundary condition $L_0=0$, the Gauss' law constraint, eq.~\eqref{eq:gauss_staggered}, leads to the solution $L_n =  a\sum_{j=1}^n Q_j$.

Notably, one can project the Hamiltonian to different charge sectors to work with states with specified quantum numbers. In the case of the Schwinger model, the projection operator $P$ to the charge neutral sector ($Q=0$) can be written as $P_{Q=0} = \sum_{n} \ket{n}\bra{n}$ where $n$ are all the parity-zero bitstrings (those with even number of 1's and 0's) due to special diagonal structure of the charge operator $Q$. On the qubits, $P_{Q=0}$ can be always written as the sum of Pauli-$Z$ operators such that $H_{Q=0}=P_{Q=0} H P_{Q=0}^\dagger$ is the projected Hamiltonian to the $Q=0$ sector\footnote{For example, at $N=4$, the projection operator is
\begin{align}
    P_{Q=0} = P_{Q=0}^\dagger = 0.375 (1 + Z_1 Z_2 Z_3 Z_4) - 0.125 \big( Z_3 Z_4 + Z_2 Z_4 + Z_2 Z_3 + Z_1 Z_4 + Z_1 Z_3 + Z_1 Z_2 \big) \nonumber
\end{align}}. Here the coefficients are determined in such a way that $P_{Q=0}^2=P_{Q=0}$. For the numerical simulation results presented in this paper, we always work in the $Q=0$ sector. This is because 
the Schwinger model spectrum does not have charged states due to confinement at all scales.
Therefore, when we say Hamiltonian, in what follows, we always mean the charge neutral Hamiltonian $H=H_{Q=0}$. 

\section{\label{sec:QITE}Quantum simulation of QED$_2$ at finite temperature}
\subsection{Quantum imaginary time evolution}
Quantum imaginary time evolution (QITE) is an efficient way to investigate eigenstates and thermal states on a quantum computer~\cite{Motta2020}. Here, we briefly review the QITE algorithm used in this work for preparing the thermal state and extracting the finite temperature observable.

In QITE, one attempts to approximate the action of the imaginary time evolution
operator on a state $| \Psi \rangle$ with a parameterized unitary
operation:
\begin{equation}
c(\Delta \beta)^{-1/2} e^{-\Delta \beta \hat{H}} | \Psi \rangle \approx e^{-i \Delta \beta \hat{A}(\vec{a})}| \Psi \rangle,
\end{equation}
where $\Delta \beta$ is a small time step and $c(\Delta \beta)^{-1/2}$ is a normalization coefficient approximated
by $1-2\Delta \beta \langle \Psi | \hat{H} | \Psi \rangle$. The real-time evolution operator $\hat{A}(\vec{a})$
is given by a linear combination of $N_\mu$ Pauli string:
\begin{align}
    \hat{A}(\vec{a}) = \sum_\mu^{N_\mu} a_\mu \hat{P}_\mu.
\end{align}
Here, $\hat{P}_{\mu}=\prod_{l}{\sigma}_{\mu_l,l}$ is a Pauli string and the subscript $\mu$ of $a_{\mu}$ labels the various Pauli strings, and their coefficients $a_\mu$ are solved from the linear equation $({\boldsymbol S}+{\boldsymbol S}^T)\,{\boldsymbol a}={\boldsymbol b}$, where the matrix ${\boldsymbol S}$ and vector ${\boldsymbol b}$ are defined by
\begin{align}
S_{\mu\nu}&=\langle{\Psi(\beta)}|\hat{P}_\nu^\dagger\hat{P}_\mu|{\Psi(\beta)}\rangle\,,\\
b_\mu&=-\frac{i}{\sqrt{c(\Delta\beta)}}\langle{\Psi(\beta)}| \big(\hat{H}\hat{P}_\mu+\hat{P}_\mu^\dagger \hat{H}\big)|{\Psi(\beta)}\rangle\,.
\end{align}
The QITE procedure is repeated for a specified
number of time steps to reach a target total evolution time. Using this iterative approach, we are able to evolve an initial quantum state $\ket{\Psi_k(0)}$ under the unitary operator $e^{-i\Delta\beta A}$ to any imaginary time $\beta$ by trotterization:
\begin{align}
|{\Psi_k(0)}\rangle \rightarrow |{\Psi_k(\beta)}\rangle +\mathcal{O}(\Delta\beta)\,.
\end{align}
Note the sum of $\ket{\Psi_k(\beta)}$ over all basis states approximates $e^{-\beta \hat{H}}/\Tr(e^{-\beta \hat{H}})$ at finite temperature $T=1/\beta$ in the sense that it gives the thermal expectation of an observable $\hat{O}$:
\begin{equation}
\frac{1}{Z}\sum_k\bra{\Psi_k(\beta/2)}\hat{O}\ket{\Psi_k(\beta/2)}\approx\langle \hat{O}\rangle_\beta=\frac{1}{Z}\Tr (e^{-\beta \hat{H}}\hat{O}),
\end{equation}
where $Z={\Tr (e^{-\beta \hat{H}})}=\sum_k\bra{\Psi_k(\beta/2)}\ket{\Psi_k(\beta/2)}$ is the canonical partition function, which is also the sum of the normalizations over all the quantum states.

\subsection{Real-time evolution at finite temperature}

Studying the dynamics of the quantum system at finite temperatures is usually complicated. In this work, we are interested in the real-time evolution of the thermal mixed state at finite temperatures of the Schwinger Hamiltonian. Using another evolution Hamiltonian $H'$ that is not equal to the Hamiltonian $H$ of the thermal system, we define a quench process by the real-time evolution of the operator. See eq.~\eqref{eq:real_time_evolution} for our quench protocol.

To implement this protocol on the quantum circuit, we first use the QITE algorithm to obtain the collection of the $\ket{\Psi_k(\beta/2)}$ states to represent the thermal state at finite temperature $T=1/\beta$. Then, we simply evolve each $\ket{\Psi_k(\beta/2)}$ in real time as usual:
\begin{equation}\label{eq:quantum_evolve}
\ket{\Psi_k(t,\beta/2)}=\mathcal{T}[e^{-i\int_0^tdt'\hat{H}'(t')}]\ket{\Psi_k(0,\beta/2)}.
\end{equation}
The final thermal expectation at real-time $t$ is
\begin{align}\label{eq:quantum_obs}
\langle \hat{O}(t)\rangle_\beta=\frac{1}{Z}\sum_k\bra{\Psi_k(t,\beta/2)}\hat{O}\ket{\Psi_k(t,\beta/2)}.   
\end{align}
Compared with eq.~\eqref{eq:real_time_evolution}, we see that thermal expectation evaluated on the quantum circuit is associated with the states instead of observables. Nevertheless, the two methods are entirely equivalent, which can be seen in our comparison between the quantum simulation and exact diagonalization results. To perform the real-time evolution, we trotterize the time into small time steps using second-order Suzuki-Trotter product formula~\cite{Berry:2005yrf, Hatano:2005gh}, which gives good convergence.

\section{\label{sec:results}Numerical results}

\subsection{Thermal state preparation}
To study the real-time evolution of physical properties at finite temperatures, we must first prepare the initial thermal state at the desired temperatures. In this work, we use the QITE algorithm developed using {\tt Qiskit}~\cite{Qiskit} in a proceeding work~\cite{Qian:2024xnr} for the thermal medium preparation. Specifically, the unitary evolution is implemented using {\tt PauliEvolutionGate} library with {\tt SuzukiTrotter} method, and the thermal expectation is evaluated with the {\tt Sampler} class on the statevector quantum simulators. In particular, we use the QITE to prepare thermal states with four qubits from $\beta/a=0$ ($aT=\infty$) to $\beta/a=4.0$ ($aT=0.25$) for the three masses $ma=0.25, 0.5, 4.0$, and at coupling $g=a^{-1}$. In particular, the thermal state is prepared from 16 basis states and a universal trotter step size $\Delta \beta = 0.01$ is used.

\begin{figure}[t]
    \centering
    \includegraphics[width=0.48\linewidth]{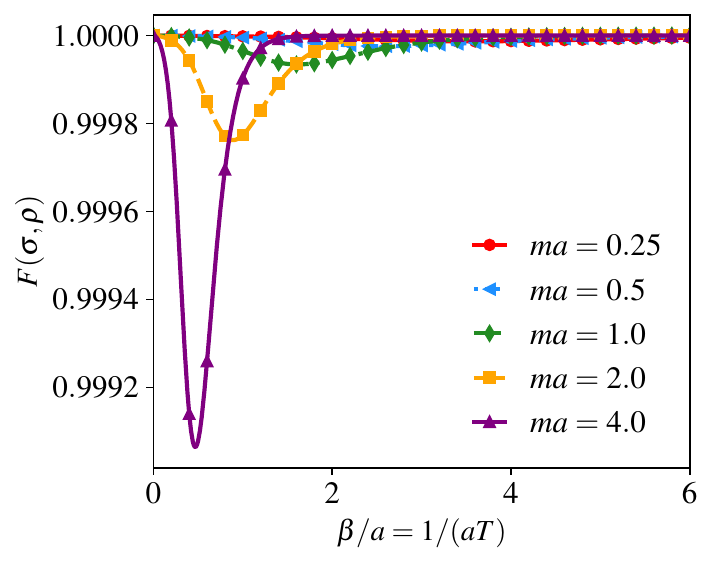}
    \includegraphics[width=0.48\linewidth]{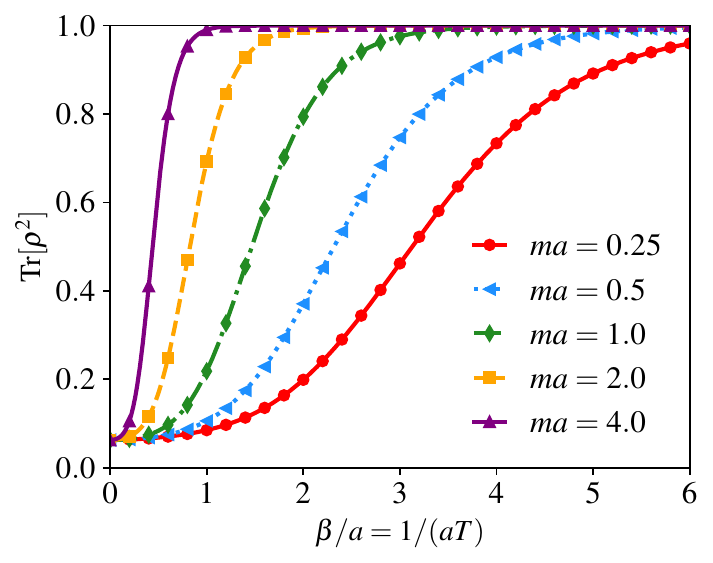}
    \caption{The fidelity and purity of the thermal state using quantum imaginary time evolution for different masses. $\Delta \beta/a=0.01$ is used for all the QITE evolution.}
    \label{fig:fidelity}
\end{figure}

To measure the validity of the quantum simulation, we use the fidelity between two density matrices $\sigma$ and $\rho$:
\begin{equation}
F(\sigma,\rho)=\left(\Tr\sqrt{\sqrt{\sigma}\rho\sqrt{\sigma}}\right)^2. 
\end{equation}
Here, we take $\sigma=e^{-\beta H}/Z$ to be the exact density matrix calculated via exact diagonalization and $\rho=\sum_k\ket{\Psi_k(\beta/2)}\bra{\Psi_k(\beta/2)}/Z$ for the thermal mixed state prepared by the QITE algorithm. In Fig.~\ref{fig:fidelity}(left), we show the fidelity of these two density matrices as a function of the QITE imaginary evolution parameter $\beta$ (i.e., inverse temperature). 
For all three cases of different masses, the fidelity is between 0.999 and 1.0, very close to 1, suggesting that the thermal state prepared by QITE is a good practical approximation of the corresponding thermal equilibrium state. In particular, we see a comparatively slight decrease of fidelity in the early $\beta$ (high temperature) region for the heaviest mass $ma=4.0$, which is due to the insufficiently small step size $\Delta \beta/a = 0.01$ used in the simulation for the heavy mass.

In Fig.~\ref{fig:fidelity}(right), we show the purity of the mixed quantum state $\rho$ at different temperatures and masses. Purity is defined by $\tr[\rho^2]$ and is an index used to determine how much a quantum state is mixed, with values equal 1 indicating a pure state, and values approaching 0 signifying a mixed state. For every mass, we expect that the purity to approach 1 as the temperature approaches 0 (equivalently, as $\beta$ approaches infinity), since the state is pure at zero temperature (i.e., the ground state). For smaller mass values, such as $ma = 0.25$, the curve shows that the purity decreases more slowly with temperature, indicating smaller mass states transition to mixed states more gently as the temperature rises. In constrast, as the mass $ma$ increases, the purity drops more rapidly. Despite these differences, all curves will eventually approach a lower purity limit as the temperature continues to increase, showing the natural tendency of quantum systems to shift towards mixed states under a finite temperature. 

It is clear that the QITE simulation with simulator provides good agreement with exact results. In principle, the procedure to simulate thermal state is the same on a real device, as demonstrated in various Ising models~\cite{Motta2020}. Nevertheless, to extract physical meaningful results, error correction and error mitigation techniques are very much needed to on today's NISQ devices.

\subsection{Real-time evolution of the charge and the current at finite temperature}

Now that the thermal state is prepared using the QITE protocol, we study real-time evolution of the total electric and axial charge densities for the system at a selected temperature with both light and heavy masses. Our simulations are conducted using the exact statevector quantum simulator provided by {\tt Qiskit}~\cite{Qiskit} and compared to the $T = 0$ limit obtained from exact diagonalization. Specifically, we perform our simulation using eqs.~\eqref{eq:quantum_evolve} and \eqref{eq:quantum_obs}, following the quench protocol provided in Sec.~\ref{sec:quench} at various chiral chemical potentials.

The real-time evolution is compared against the zero-temperature limit using exact diagonalization, serving as a benchmark for the results obtained from the quantum simulator at finite temperatures. We first note that the results at $T=0$ are consistent with the results in~\cite{Kharzeev:2020kgc}, which performed the study under the periodic boundary condition at $g=0$ while we work with the open boundary condition at $g\neq0$. We also observe that $\langle Q\rangle =0$ at all finite temperatures and it does not evolve over time, which is expected, since $[H(t),Q]=0$ at all time $t$. The quantum simulation uses four qubits and applies the QITE algorithm for thermal state preparation. These methods allow for accurate modeling of thermal states and real-time evolution, offering a robust framework for understanding the underlying physics. The second order Suzuki-Trotter product formula~\cite{Berry:2005yrf, Hatano:2005gh} is used for the real-time evolution, with step sizes $dt$ varying based on the masses: 0.1 for light masses and 0.01 for heavy masses. 

\begin{figure}[t]
    \centering
    \subfigure[$ma = 0.25$]{\includegraphics[width=0.98\textwidth]{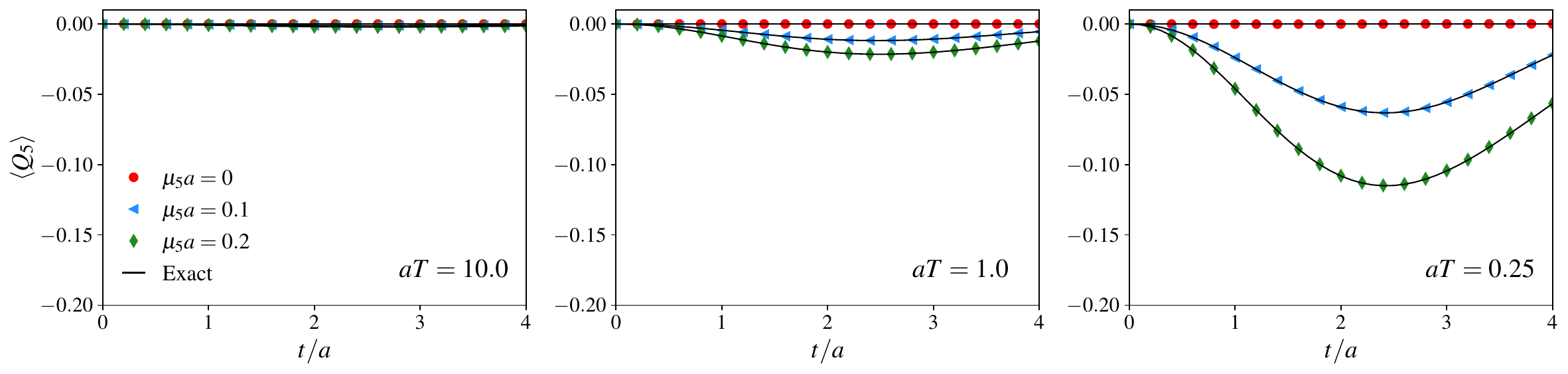}}
    \subfigure[$ma = 1.0$]{\includegraphics[width=0.98\textwidth]{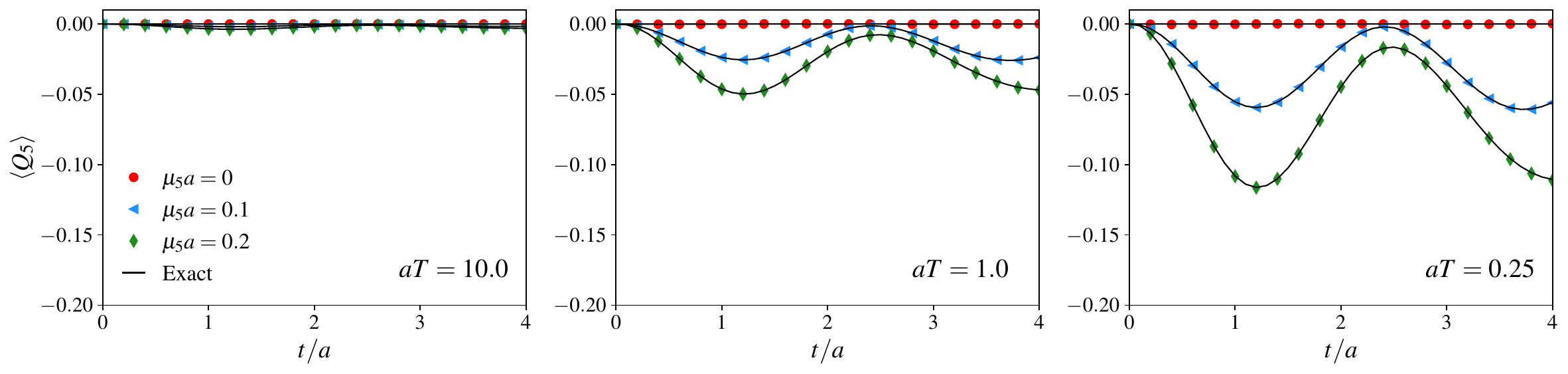}}
    \subfigure[$ma = 4.0$]{\includegraphics[width=0.98\textwidth]{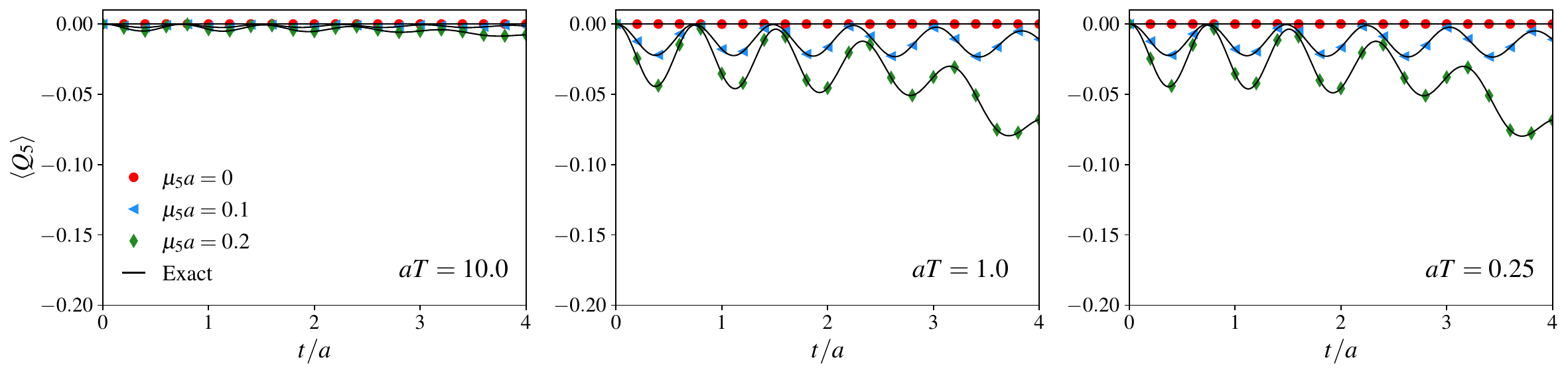}}
    \caption{Real-time evolution of the total axial charge densities at finite temperatures using quantum simulator compared with exact diagonalization. Quantum simulation results using $N=4$ qubits are in colored markers; exact diagonalization results are in solid lines. Specifically, we use the QITE algorithm for the thermal state and the second-order Suzuki Trotter formula for the time evolution with step size $dt/a=0.1, 0.1, 0.02$ for the three masses, respectively.}
    \label{fig:axial_charge_N4_with_exact}
\end{figure}

Fig.~\ref{fig:axial_charge_N4_with_exact} shows the real-time evolution of the thermal average of the axial charge $\langle Q_5(t) \rangle_\beta$ for various temperatures, $aT = 10.0,1.0,0.25$. The mass takes values of $ma=0.25, 1.0, 4.0$ to examine how light and heavy masses affect the evolution of the vector and axial charge densities, showing how it evolves under different temperatures. Similar to the state preparation, the thermal expectation is also evaluated from 16 basis states in the simulation with $N=4$ qubits. The results highlight the role of a chiral chemical potential $\mu_5$ and include both exact solutions and approximate results using quantum simulation. To confirm consistency with theoretical expectations, we show that the total vector charge is conserved and does not depend on time. 

In studying the chiral dynamics in the Schwinger model, our results share several important similarities with our previous results of nonlinear chiral magnetic waves (CMWs)~\cite{Ikeda:2023vfk}, particularly in observing how the oscillation of axial charges is influenced by mass. Here, chiral magnetic waves are collective excitations in a (1+1)-dimensional chiral medium under the influence of a magnetic field. These waves are a direct consequence of the interplay between axial (chiral) and vector (electric) currents in the presence of magnetic fields. Notably, the tendency that heavier mass results in faster oscillations of the axial charge is common~\cite{PhysRevD.83.085007}.
In both studies, it is observed that the axial charge density oscillates more rapidly with increasing mass. In the nonlinear CMWs, the nonlinearity arises because, in massive Schwinger model, the frequency of axial charge oscillations becomes significantly higher than that of the vector charge. This results in the axial charge oscillating rapidly while being nearly confined within static electric dipoles, a phenomenon referred to as the ``thumper'' solution ~\cite{Ikeda:2023vfk}. The interplay between the excited state (causing rapid oscillations) and the ground state (with almost static behavior) creates a non-linear dynamic in the system.

\begin{figure}
    \centering
    \subfigure[$ma = 1.0$]{
    \includegraphics[width=0.43\textwidth]{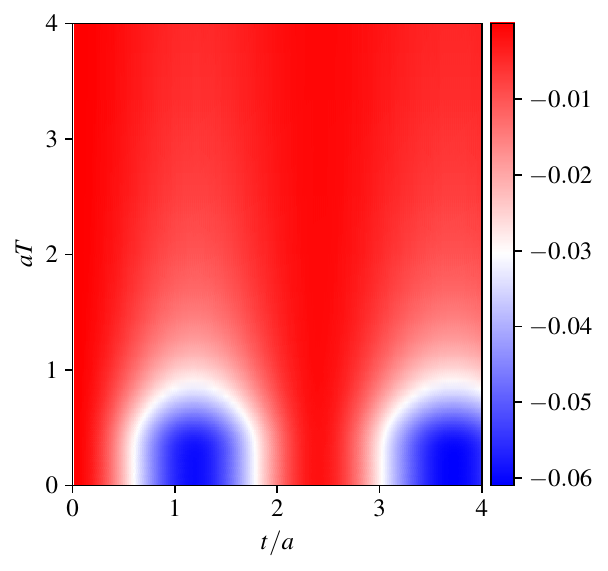}}
    \subfigure[$ma = 4.0$]{
    \includegraphics[width=0.43\textwidth]{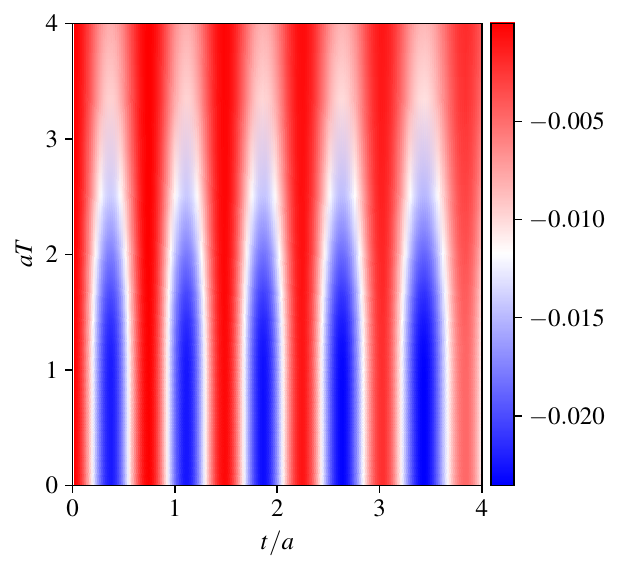}}
    \caption{Time evolution of the axial charge density using quantum simulator at various temperatures on a lattice with $N=4$ qubits for (a) the light and (b) the heavy masses. Chiral chemical potential $\mu_5 a = 0.1$ is used. Interpolation is used to present the simulation results in the $aT$ axis.}
    \label{fig:heatmap_N4}
\end{figure}

In Fig.~\ref{fig:axial_charge_N4_with_exact}, we present a comparison between our simulation results and exact diagonalization, observing good agreement when sufficiently small time steps are used. This figure illustrates how accurately our simulation can predict the system's behavior, validating our approach. Furthermore, the rapid oscillations observed in the simulation at a larger mass $(ma = 4)$ also verified our expectation. These oscillations are attributed to the interplay between different energy eigenstates, and their amplitude increases as the temperature rises. The consistency between the simulation in Fig.~\ref{fig:axial_charge_N4_with_exact} underscores the accuracy of our model in capturing the dynamic properties of the system at varying conditions.

To study the dynamical interplay between finite temperature and real-time evolutions, we also present the density plots for $\langle Q_5(t) \rangle_\beta$ in Fig.~\ref{fig:heatmap_N4}. The axial charge oscillates as a function of time $t$ at different temperatures $T$, with increasing frequency for the heavier mass. Notably, the thermal state is first obtained using the QITE algorithm at finite $\beta/a=1/(aT)$ values in the range of [0, 4], and then we obtain the density plot versus $aT$ to the range of $[\infty, 0.25]$ with interpolation to $T=0$ temperature. The number of modes for the time oscillations is directly related to the mass in the Schwinger Hamiltonian. 

\begin{figure}
    \centering
    \subfigure[$m/a = 1.0\, (a=1.0\, \mathrm{GeV}^{-1})$]{\includegraphics[width=0.45\textwidth]{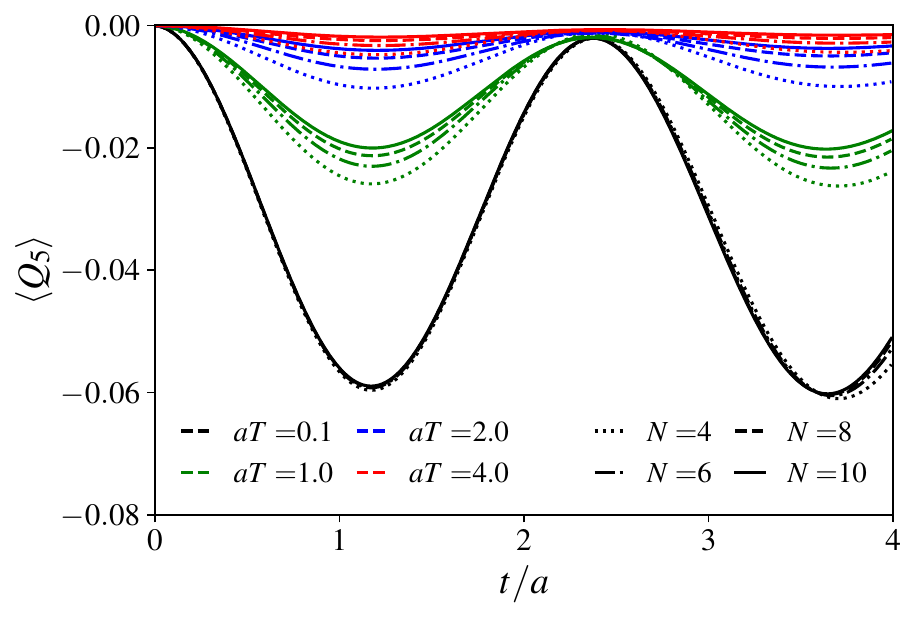}}\qquad
    \subfigure[$m/a = 4.0\, (a=1.0\, \mathrm{GeV}^{-1})$]{\includegraphics[width=0.45\textwidth]{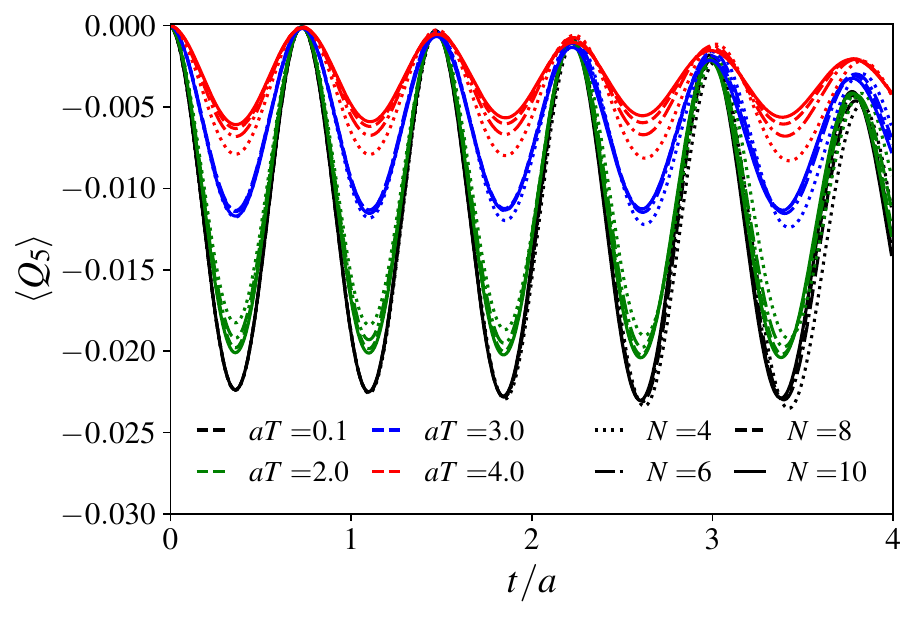}}
    \subfigure[$m/a = 1.0\, (a=0.5\, \mathrm{GeV}^{-1})$]{\includegraphics[width=0.45\textwidth]{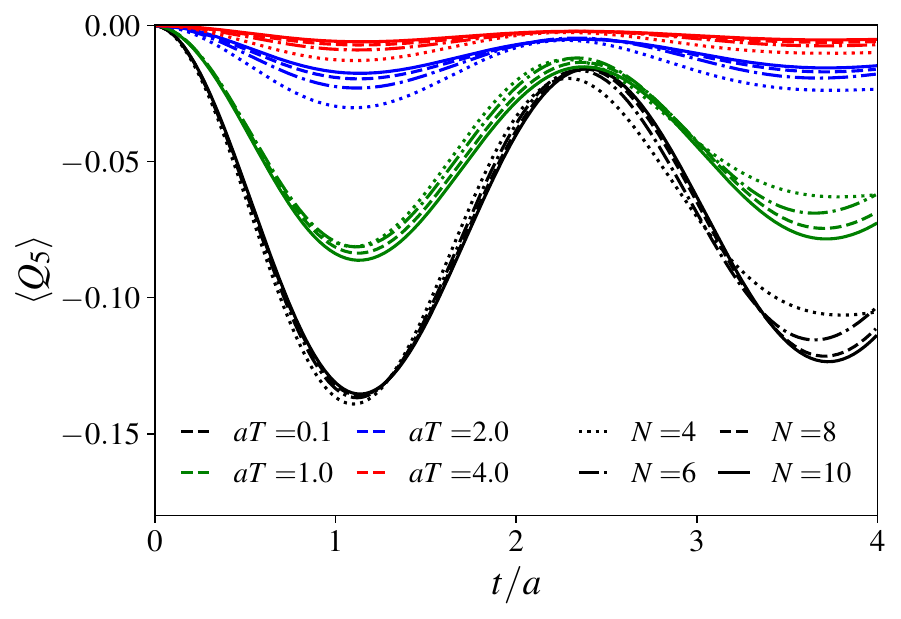}}\qquad
    \subfigure[$m/a = 4.0\, (a=0.5\, \mathrm{GeV}^{-1})$]{\includegraphics[width=0.45\textwidth]{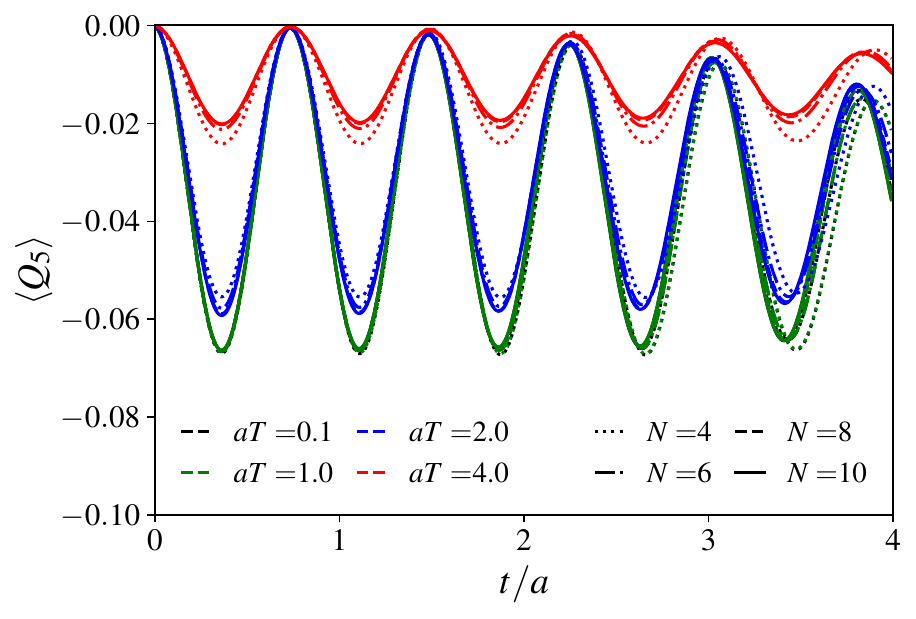}}
    \caption{Time evolution of the axial charge density at selected temperature $T$ using exact diagonalization on increasing lattice sizes. Lattice/qubit sizes of $N=4,6,8,10$ and two lattice spacings $a=1.0\, \mathrm{GeV}^{-1}$ and $0.5\, \mathrm{GeV}^{-1}$ are used. Chiral chemical potential $\mu_5 a = 0.1$ is used.
    }
    \label{fig:increasing_N}
\end{figure}

With the goal of extrapolating our results in the continuum, we perform the exact diagonalization at an increasing number of qubits from $N=4$ to $N=10$. Importantly, we have direct access to the finite temperature thermal state by using exact matrix exponentiation for the trotterized small time steps following eq.~\eqref{eq:thermal_observable}. In Fig.~\ref{fig:increasing_N}, we present the time evolution of the axial charge densities at selected temperatures for an increasing number of qubits for two selected lattice spacings $a=1.0\,\mathrm{GeV}^{-1}$ and $a=0.5\,\mathrm{GeV}^{-1}$. As we can see, the results appear to converge nicely with increasing $N$, especially in the low-temperature region, which makes the $N=10$ simulation result a good representation for a large system. Furthermore, the simulation results at two different lattice spacings show qualitatively the same pattern. Nonetheless, comprehensive analysis is still needed for extrapolation to a continuum theory, which we leave for future work.

\begin{figure}
    \centering
    \subfigure[$ma = 0.25$]{\includegraphics[width=0.42\textwidth]{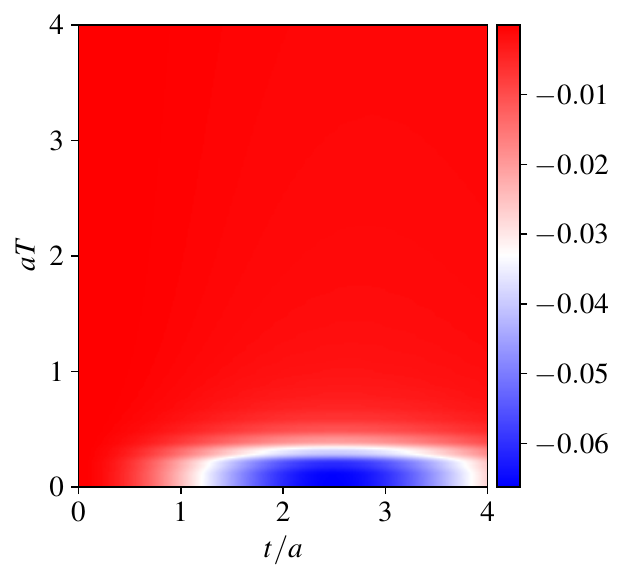}}\qquad
    \subfigure[$ma = 1.0$]{\includegraphics[width=0.42\textwidth]{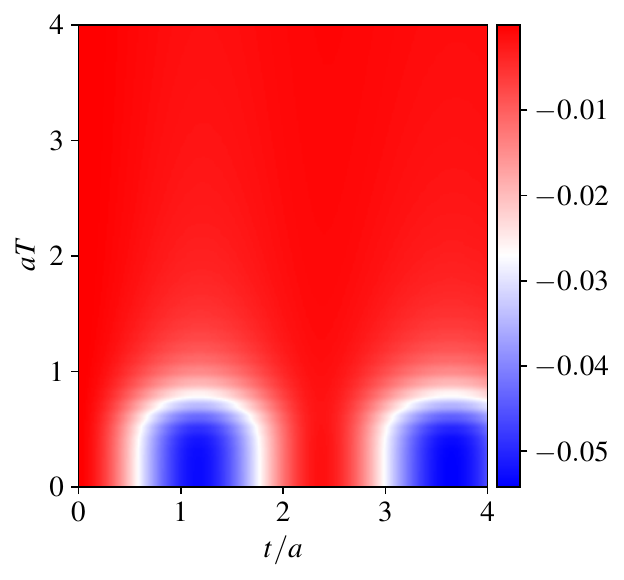}}
    \subfigure[$ma = 2.0$]{\includegraphics[width=0.42\textwidth]{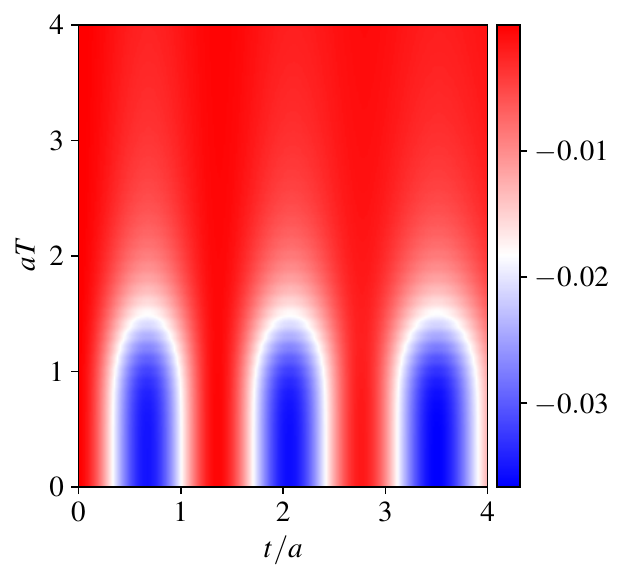}}\qquad
    \subfigure[$ma = 4.0$]{\includegraphics[width=0.42\textwidth]{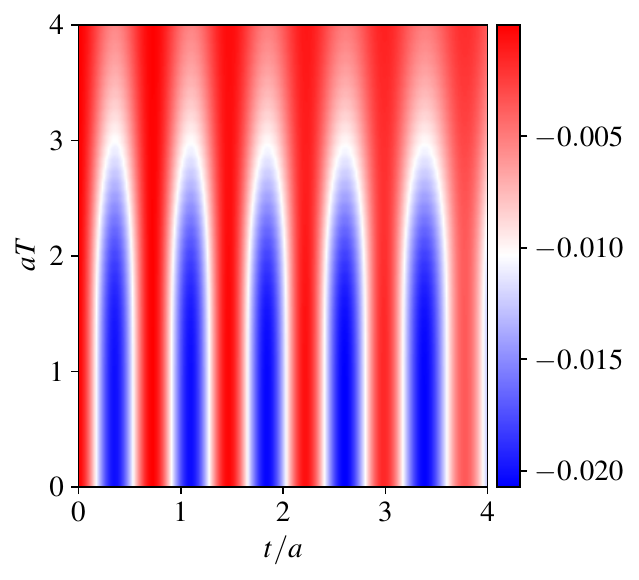}}
    \caption{Time evolution of the axial charge density at various temperature $T$ using exact diagonalization on a lattice with qubits $N=10$. Chiral chemical potential $\mu_5 a = 0.1$ is used. 
    }
    \label{fig:heatmap_N10}
\end{figure}

In Fig.~\ref{fig:heatmap_N10}, we show the exact results of the density plot of the thermal expectation value of $\langle Q_5(t)\rangle_T$ in terms of $aT$ and $(t/a)$ (with temperature in units of $aT$ in the range of $[0, 4]$). Comparing to results simulated on the quantum simulator shown in Fig.~\ref{fig:heatmap_N4}, we observe the same behavior for both $ma=1.0$ and $4.0$ cases, despite having many more qubits. Quantitatively, we can also observe good agreement between the quantum and classical simulations by counting the oscillation modes and the peak intensity values in the density plots.

\begin{figure}
    \centering
    \subfigure[$ma = 1.0$]{\includegraphics[width=0.42\textwidth]{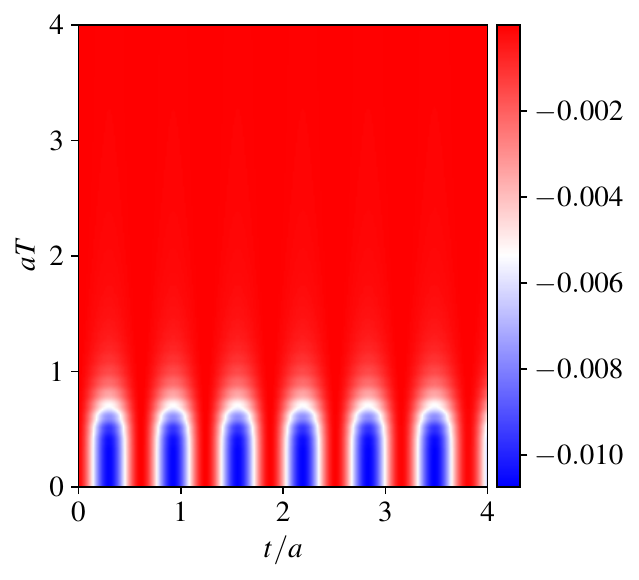}}\qquad
    \subfigure[$ma = 4.0$]{\includegraphics[width=0.42\textwidth]{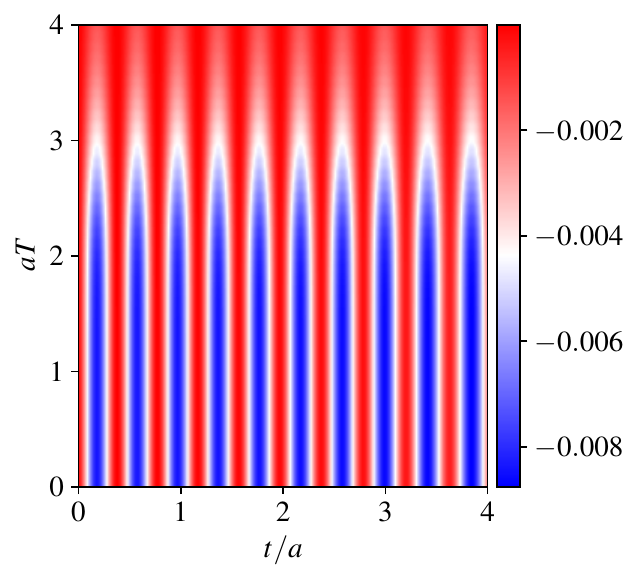}}
    \caption{Time evolution of the axial charge density at strong coupling $g=4a^{-1}$ at various temperature $T$ using exact diagonalization on a lattice with qubits $N=10$. Chiral chemical potential $\mu_5 a = 0.1$ is used. 
    }
    \label{fig:heatmap_N10_g4}
\end{figure}

It is also interesting to investigate the charge dependence on the coupling $g$. Throughout the paper, we have fixed $g=a^{-1}$. In Fig.~\ref{fig:heatmap_N10_g4}, we present the axial charge density evolution at a larger coupling $g=4a^{-1}$ using exact diagonalization with $N=10$ qubits. Compared to our previous results of the same masses, we can see that the oscillation becomes a lot faster with the increased coupling, but the thermal damping sets in at about the same temperature. This is because the interactions are controlled by the fermion mass, not by the values of the coupling.

\section{\label{sec:conclusion}Conclusion}
This paper investigates the real-time dynamics of the chiral magnetic effect (CME) at finite temperature in the context of the massive Schwinger model, quantum electrodynamics (QED) in $(1+1)$ dimensions. The study utilizes quantum simulation techniques, combining the quantum imaginary time evolution (QITE) algorithm for thermal state preparation and trotterization for real-time evolution, to examine the thermal properties of the system and the transitions induced by a chiral chemical potential via a quench process.

Our results illustrate the dynamic interplay between the fermion mass and the real-time dynamics of the axial charge. At zero temperature, the corresponding dynamics has been investigated in our previous studies  ~\cite{Kharzeev:2020kgc,Ikeda:2023vfk,Czajka2022}. 
The new result in this paper is the temperature dependence of the real-time CME response. Specifically, at temperatures small compared to the fermion mass, the CME response is similar to the one found previously at zero temperature. However, at larger temperatures, the oscillations of the axial charge induced by the CME are dampened by thermal effects. This can be understood as a result of the contribution from excited fermion modes to the real-time CME response, that may depend on the temperature \cite{PhysRevD.80.034028}. 

In addition, at large fermion mass the axial charge oscillations become very non-linear. The origin of this  non-linearity can be explained in the following way: in bosonized description of the massive Schwinger model, fermion mass induces non-linear interactions among the bosons, so the chiral magnetic waves become highly non-linear \cite{Ikeda:2023vfk}.

Finally, our study of CME dynamics suggests that quantum simulations can be used to study other real-time phenomena at finite temperatures in gauge theories that form the basis of the Standard Model.

\section*{Acknowledgement}
We thank Meijian Li for helpful comment on the manuscript. We are also grateful to Andrea Palermo and 
Ismail Zahed for useful discussions. This work was supported by the U.S. Department of Energy, Office of Science, National Quantum Information Science Research Centers, Co-design Center for Quantum Advantage (C2QA) under Contract No.DE-SC0012704 (KI, DK, FZ), the U.S. Department of Energy, Office of Science, Office of Nuclear Physics, Grants Nos. DE-FG88ER41450 (DK), DE-SC0012704 (DK) and DE-SC0011090 (FZ), and the  National Science Foundation under grant No. PHY-1945471 (ZK). WQ is supported by the European Research Council under project ERC-2018-ADG-835105 YoctoLHC; by the Spanish Research State Agency under project PID2020-119632GB-I00; by Xunta de Galicia (Centro singular de investigacion de Galicia accreditation 2019-2022), by European Union ERDF, and by the Marie Sklodowska-Curie Actions Postdoctoral Fellowships under Grant No. 101109293.

\bibliographystyle{JHEP}
\bibliography{main.bbl}

\providecommand{\href}[2]{#2}\begingroup\raggedright\begin{thebibliography}{10}

\bibitem{Jordan:2011ci}
S.P.~Jordan, K.S.M.~Lee and J.~Preskill, \emph{{Quantum Computation of
  Scattering in Scalar Quantum Field Theories}}, {\emph{Quant. Inf. Comput.}
  {\bfseries 14} (2014) 1014}
  [\href{https://arxiv.org/abs/1112.4833}{{\ttfamily 1112.4833}}].

\bibitem{Jordan:2012xnu}
S.P.~Jordan, K.S.M.~Lee and J.~Preskill, \emph{{Quantum Algorithms for Quantum
  Field Theories}},
  \href{https://doi.org/10.1126/science.1217069}{\emph{Science} {\bfseries 336}
  (2012) 1130} [\href{https://arxiv.org/abs/1111.3633}{{\ttfamily 1111.3633}}].

\bibitem{Jordan:2014tma}
S.P.~Jordan, K.S.M.~Lee and J.~Preskill, \emph{{Quantum Algorithms for
  Fermionic Quantum Field Theories}},
  \href{https://arxiv.org/abs/1404.7115}{{\ttfamily 1404.7115}}.

\bibitem{Bauer:2022hpo}
C.W.~Bauer, Z.~Davoudi, A.B.~Balantekin, T.~Bhattacharya, M.~Carena,
  W.A.~de~Jong et~al., \emph{Quantum simulation for high-energy physics},
  \href{https://doi.org/10.1103/PRXQuantum.4.027001}{\emph{PRX Quantum}
  {\bfseries 4} (2023) 027001}.

\bibitem{Bauer:2023qgm}
C.W.~Bauer, Z.~Davoudi, N.~Klco and M.J.~Savage, \emph{{Quantum simulation of
  fundamental particles and forces}},
  \href{https://doi.org/10.1038/s42254-023-00599-8}{\emph{Nature Rev. Phys.}
  {\bfseries 5} (2023) 420} [\href{https://arxiv.org/abs/2404.06298}{{\ttfamily
  2404.06298}}].

\bibitem{Czajka2022}
A.M.~Czajka, Z.-B.~Kang, H.~Ma and F.~Zhao, \emph{Quantum simulation of chiral
  phase transitions},
  \href{https://doi.org/10.1007/JHEP08(2022)209}{\emph{Journal of High Energy
  Physics} {\bfseries 2022} (2022) 209}.

\bibitem{Czajka:2022plx}
A.M.~Czajka, Z.-B.~Kang, Y.~Tee and F.~Zhao, \emph{{Studying chirality
  imbalance with quantum algorithms}},
  \href{https://arxiv.org/abs/2210.03062}{{\ttfamily 2210.03062}}.

\bibitem{Davoudi:2022uzo}
Z.~Davoudi, N.~Mueller and C.~Powers, \emph{Towards quantum computing phase
  diagrams of gauge theories with thermal pure quantum states},
  \href{https://doi.org/10.1103/PhysRevLett.131.081901}{\emph{Phys. Rev. Lett.}
  {\bfseries 131} (2023) 081901}.

\bibitem{Qian:2024xnr}
W.~Qian and B.~Wu, \emph{{Quantum computation in fermionic thermal field
  theories}}, \href{https://doi.org/10.1007/JHEP07(2024)166}{\emph{JHEP}
  {\bfseries 07} (2024) 166}
  [\href{https://arxiv.org/abs/2404.07912}{{\ttfamily 2404.07912}}].

\bibitem{Kreshchuk:2020dla}
M.~Kreshchuk, W.M.~Kirby, G.~Goldstein, H.~Beauchemin and P.J.~Love,
  \emph{{Quantum simulation of quantum field theory in the light-front
  formulation}}, \href{https://doi.org/10.1103/PhysRevA.105.032418}{\emph{Phys.
  Rev. A} {\bfseries 105} (2022) 032418}
  [\href{https://arxiv.org/abs/2002.04016}{{\ttfamily 2002.04016}}].

\bibitem{Qian:2021jxp}
W.~Qian, R.~Basili, S.~Pal, G.~Luecke and J.P.~Vary, \emph{{Solving hadron
  structures using the basis light-front quantization approach on quantum
  computers}},
  \href{https://doi.org/10.1103/PhysRevResearch.4.043193}{\emph{Phys. Rev.
  Res.} {\bfseries 4} (2022) 043193}
  [\href{https://arxiv.org/abs/2112.01927}{{\ttfamily 2112.01927}}].

\bibitem{Barata:2023clv}
J.a.~Barata, X.~Du, M.~Li, W.~Qian and C.A.~Salgado, \emph{{Quantum simulation
  of in-medium QCD jets: Momentum broadening, gluon production, and entropy
  growth}}, \href{https://doi.org/10.1103/PhysRevD.108.056023}{\emph{Phys. Rev.
  D} {\bfseries 108} (2023) 056023}
  [\href{https://arxiv.org/abs/2307.01792}{{\ttfamily 2307.01792}}].

\bibitem{Du:2023bpw}
W.~Du and J.P.~Vary, \emph{{Multinucleon structure and dynamics via quantum
  computing}}, \href{https://doi.org/10.1103/PhysRevA.108.052614}{\emph{Phys.
  Rev. A} {\bfseries 108} (2023) 052614}
  [\href{https://arxiv.org/abs/2304.04838}{{\ttfamily 2304.04838}}].

\bibitem{McArdle:2018tza}
S.~McArdle, S.~Endo, A.~Aspuru-Guzik, S.C.~Benjamin and X.~Yuan, \emph{{Quantum
  computational chemistry}},
  \href{https://doi.org/10.1103/revmodphys.92.015003}{\emph{Rev. Mod. Phys.}
  {\bfseries 92} (2020) 015003}
  [\href{https://arxiv.org/abs/1808.10402}{{\ttfamily 1808.10402}}].

\bibitem{Schwinger:1962tp}
J.S.~Schwinger, \emph{{Gauge Invariance and Mass. 2.}},
  \href{https://doi.org/10.1103/PhysRev.128.2425}{\emph{Phys. Rev.} {\bfseries
  128} (1962) 2425}.

\bibitem{Klco:2018kyo}
N.~Klco, E.F.~Dumitrescu, A.J.~McCaskey, T.D.~Morris, R.C.~Pooser, M.~Sanz
  et~al., \emph{{Quantum-classical computation of Schwinger model dynamics
  using quantum computers}},
  \href{https://doi.org/10.1103/PhysRevA.98.032331}{\emph{Phys. Rev. A}
  {\bfseries 98} (2018) 032331}
  [\href{https://arxiv.org/abs/1803.03326}{{\ttfamily 1803.03326}}].

\bibitem{Farrell:2023fgd}
R.C.~Farrell, M.~Illa, A.N.~Ciavarella and M.J.~Savage, \emph{{Scalable
  Circuits for Preparing Ground States on Digital Quantum Computers: The
  Schwinger Model Vacuum on 100 Qubits}},
  \href{https://doi.org/10.1103/PRXQuantum.5.020315}{\emph{PRX Quantum}
  {\bfseries 5} (2024) 020315}
  [\href{https://arxiv.org/abs/2308.04481}{{\ttfamily 2308.04481}}].

\bibitem{Farrell:2024fit}
R.C.~Farrell, M.~Illa, A.N.~Ciavarella and M.J.~Savage, \emph{{Quantum
  simulations of hadron dynamics in the Schwinger model using 112 qubits}},
  \href{https://doi.org/10.1103/PhysRevD.109.114510}{\emph{Phys. Rev. D}
  {\bfseries 109} (2024) 114510}
  [\href{https://arxiv.org/abs/2401.08044}{{\ttfamily 2401.08044}}].

\bibitem{Butt:2019uul}
N.~Butt, S.~Catterall, Y.~Meurice, R.~Sakai and J.~Unmuth-Yockey, \emph{{Tensor
  network formulation of the massless Schwinger model with staggered
  fermions}}, \href{https://doi.org/10.1103/PhysRevD.101.094509}{\emph{Phys.
  Rev. D} {\bfseries 101} (2020) 094509}
  [\href{https://arxiv.org/abs/1911.01285}{{\ttfamily 1911.01285}}].

\bibitem{Magnifico:2019kyj}
G.~Magnifico, M.~Dalmonte, P.~Facchi, S.~Pascazio, F.V.~Pepe and E.~Ercolessi,
  \emph{{Real Time Dynamics and Confinement in the $\mathbb{Z}_{n}$
  Schwinger-Weyl lattice model for 1+1 QED}},
  \href{https://doi.org/10.22331/q-2020-06-15-281}{\emph{Quantum} {\bfseries 4}
  (2020) 281} [\href{https://arxiv.org/abs/1909.04821}{{\ttfamily
  1909.04821}}].

\bibitem{Shaw:2020udc}
A.F.~Shaw, P.~Lougovski, J.R.~Stryker and N.~Wiebe, \emph{{Quantum Algorithms
  for Simulating the Lattice Schwinger Model}},
  \href{https://doi.org/10.22331/q-2020-08-10-306}{\emph{Quantum} {\bfseries 4}
  (2020) 306} [\href{https://arxiv.org/abs/2002.11146}{{\ttfamily
  2002.11146}}].

\bibitem{Kharzeev:2020kgc}
D.E.~Kharzeev and Y.~Kikuchi, \emph{{Real-time chiral dynamics from a digital
  quantum simulation}},
  \href{https://doi.org/10.1103/PhysRevResearch.2.023342}{\emph{Phys. Rev.
  Res.} {\bfseries 2} (2020) 023342}
  [\href{https://arxiv.org/abs/2001.00698}{{\ttfamily 2001.00698}}].

\bibitem{Ikeda:2020agk}
K.~Ikeda, D.E.~Kharzeev and Y.~Kikuchi, \emph{{Real-time dynamics of
  Chern-Simons fluctuations near a critical point}},
  \href{https://doi.org/10.1103/PhysRevD.103.L071502}{\emph{Phys. Rev. D}
  {\bfseries 103} (2021) L071502}
  [\href{https://arxiv.org/abs/2012.02926}{{\ttfamily 2012.02926}}].

\bibitem{2023arXiv230111991F}
A.~Florio, D.~Frenklakh, K.~Ikeda, D.~Kharzeev, V.~Korepin, S.~Shi et~al.,
  \emph{Real-time nonperturbative dynamics of jet production in schwinger
  model: Quantum entanglement and vacuum modification},
  \href{https://doi.org/10.1103/PhysRevLett.131.021902}{\emph{Phys. Rev. Lett.}
  {\bfseries 131} (2023) 021902}.

\bibitem{PhysRevD.107.L071502}
K.~Ikeda, \emph{Criticality of quantum energy teleportation at phase transition
  points in quantum field theory},
  \href{https://doi.org/10.1103/PhysRevD.107.L071502}{\emph{Phys. Rev. D}
  {\bfseries 107} (2023) L071502}.

\bibitem{Ikeda:2023zil}
K.~Ikeda, D.E.~Kharzeev, R.~Meyer and S.~Shi, \emph{Detecting the critical
  point through entanglement in the schwinger model},
  \href{https://doi.org/10.1103/PhysRevD.108.L091501}{\emph{Phys. Rev. D}
  {\bfseries 108} (2023) L091501}.

\bibitem{Ikeda:2023vfk}
K.~Ikeda, D.E.~Kharzeev and S.~Shi, \emph{Nonlinear chiral magnetic waves},
  \href{https://doi.org/10.1103/PhysRevD.108.074001}{\emph{Phys. Rev. D}
  {\bfseries 108} (2023) 074001}.

\bibitem{Florio:2024aix}
A.~Florio, D.~Frenklakh, K.~Ikeda, D.E.~Kharzeev, V.~Korepin, S.~Shi et~al.,
  \emph{{Quantum simulation of entanglement and hadronization in jet
  production: lessons from the massive Schwinger model}},
  \href{https://arxiv.org/abs/2404.00087}{{\ttfamily 2404.00087}}.

\bibitem{Kharzeev:2004ey}
D.~Kharzeev, \emph{{Parity violation in hot QCD: Why it can happen, and how to
  look for it}},
  \href{https://doi.org/10.1016/j.physletb.2005.11.075}{\emph{Phys. Lett. B}
  {\bfseries 633} (2006) 260}
  [\href{https://arxiv.org/abs/hep-ph/0406125}{{\ttfamily hep-ph/0406125}}].

\bibitem{2008NuPhA.803..227K}
D.E.~Kharzeev, L.D.~McLerran and H.J.~Warringa, \emph{The effects of
  topological charge change in heavy ion collisions: “event by event p and cp
  violation”},
  \href{https://doi.org/https://doi.org/10.1016/j.nuclphysa.2008.02.298}{\emph{Nuclear
  Physics A} {\bfseries 803} (2008) 227}.

\bibitem{Fukushima:2008xe}
K.~Fukushima, D.E.~Kharzeev and H.J.~Warringa, \emph{{The Chiral Magnetic
  Effect}}, \href{https://doi.org/10.1103/PhysRevD.78.074033}{\emph{Phys. Rev.
  D} {\bfseries 78} (2008) 074033}
  [\href{https://arxiv.org/abs/0808.3382}{{\ttfamily 0808.3382}}].

\bibitem{Kharzeev:2013ffa}
D.E.~Kharzeev, \emph{{The Chiral Magnetic Effect and Anomaly-Induced
  Transport}}, \href{https://doi.org/10.1016/j.ppnp.2014.01.002}{\emph{Prog.
  Part. Nucl. Phys.} {\bfseries 75} (2014) 133}
  [\href{https://arxiv.org/abs/1312.3348}{{\ttfamily 1312.3348}}].

\bibitem{Kharzeev:2015znc}
D.E.~Kharzeev, J.~Liao, S.A.~Voloshin and G.~Wang, \emph{{Chiral magnetic and
  vortical effects in high-energy nuclear collisions\textemdash{}A status
  report}}, \href{https://doi.org/10.1016/j.ppnp.2016.01.001}{\emph{Prog. Part.
  Nucl. Phys.} {\bfseries 88} (2016) 1}
  [\href{https://arxiv.org/abs/1511.04050}{{\ttfamily 1511.04050}}].

\bibitem{Landsteiner:2016led}
K.~Landsteiner, \emph{{Notes on Anomaly Induced Transport}},
  \href{https://doi.org/10.5506/APhysPolB.47.2617}{\emph{Acta Phys. Polon. B}
  {\bfseries 47} (2016) 2617}
  [\href{https://arxiv.org/abs/1610.04413}{{\ttfamily 1610.04413}}].

\bibitem{Kharzeev:2020jxw}
D.E.~Kharzeev and J.~Liao, \emph{{Chiral magnetic effect reveals the topology
  of gauge fields in heavy-ion collisions}},
  \href{https://doi.org/10.1038/s42254-020-00254-6}{\emph{Nature Rev. Phys.}
  {\bfseries 3} (2021) 55} [\href{https://arxiv.org/abs/2102.06623}{{\ttfamily
  2102.06623}}].

\bibitem{2016NatPh..12..550L}
Q.~{Li}, D.E.~{Kharzeev}, C.~{Zhang}, Y.~{Huang}, I.~{Pletikosi{\'c}},
  A.V.~{Fedorov} et~al., \emph{{Chiral magnetic effect in ZrTe$_{5}$}},
  \href{https://doi.org/10.1038/nphys3648}{\emph{Nature Physics} {\bfseries 12}
  (2016) 550} [\href{https://arxiv.org/abs/1412.6543}{{\ttfamily 1412.6543}}].

\bibitem{PhysRevD.80.034028}
D.E.~Kharzeev and H.J.~Warringa, \emph{Chiral magnetic conductivity},
  \href{https://doi.org/10.1103/PhysRevD.80.034028}{\emph{Phys. Rev. D}
  {\bfseries 80} (2009) 034028}.

\bibitem{PhysRevLett.98.070201}
G.~Vidal, \emph{Classical simulation of infinite-size quantum lattice systems
  in one spatial dimension},
  \href{https://doi.org/10.1103/PhysRevLett.98.070201}{\emph{Phys. Rev. Lett.}
  {\bfseries 98} (2007) 070201}.

\bibitem{PhysRevB.78.155117}
R.~Or\'us and G.~Vidal, \emph{Infinite time-evolving block decimation algorithm
  beyond unitary evolution},
  \href{https://doi.org/10.1103/PhysRevB.78.155117}{\emph{Phys. Rev. B}
  {\bfseries 78} (2008) 155117}.

\bibitem{PhysRevB.91.115137}
H.N.~Phien, I.P.~McCulloch and G.~Vidal, \emph{Fast convergence of imaginary
  time evolution tensor network algorithms by recycling the environment},
  \href{https://doi.org/10.1103/PhysRevB.91.115137}{\emph{Phys. Rev. B}
  {\bfseries 91} (2015) 115137}.

\bibitem{Motta2020}
M.~Motta, C.~Sun, A.T.K.~Tan, M.J.~O'Rourke, E.~Ye, A.J.~Minnich et~al.,
  \emph{Determining eigenstates and thermal states on a quantum computer using
  quantum imaginary time evolution},
  \href{https://doi.org/10.1038/s41567-019-0704-4}{\emph{Nature Physics}
  {\bfseries 16} (2020) 205}.

\bibitem{Gomes:2021ckn}
N.~Gomes, A.~Mukherjee, F.~Zhang, T.~Iadecola, C.-Z.~Wang, K.-M.~Ho et~al.,
  \emph{{Adaptive Variational Quantum Imaginary Time Evolution Approach for
  Ground State Preparation}},
  \href{https://doi.org/10.1002/qute.202100114}{\emph{Adv. Quantum Technol.}
  {\bfseries 4} (2021) 2100114}
  [\href{https://arxiv.org/abs/2102.01544}{{\ttfamily 2102.01544}}].

\bibitem{PhysRevLett.111.010401}
S.~Sugiura and A.~Shimizu, \emph{Canonical thermal pure quantum state},
  \href{https://doi.org/10.1103/PhysRevLett.111.010401}{\emph{Phys. Rev. Lett.}
  {\bfseries 111} (2013) 010401}.

\bibitem{McArdle2019}
S.~{McArdle}, T.~{Jones}, S.~{Endo}, Y.~{Li}, S.C.~{Benjamin} and X.~{Yuan},
  \emph{{Variational ansatz-based quantum simulation of imaginary time
  evolution}}, \href{https://doi.org/10.1038/s41534-019-0187-2}{\emph{npj
  Quantum Information} {\bfseries 5} (2019) 75}.

\bibitem{PhysRevC.109.044322}
Y.H.~Li, J.~Al-Khalili and P.~Stevenson, \emph{Quantum simulation approach to
  implementing nuclear density functional theory via imaginary time evolution},
  \href{https://doi.org/10.1103/PhysRevC.109.044322}{\emph{Phys. Rev. C}
  {\bfseries 109} (2024) 044322}.

\bibitem{Yuan2019theoryofvariational}
X.~Yuan, S.~Endo, Q.~Zhao, Y.~Li and S.C.~Benjamin, \emph{Theory of variational
  quantum simulation},
  \href{https://doi.org/10.22331/q-2019-10-07-191}{\emph{{Quantum}} {\bfseries
  3} (2019) 191}.

\bibitem{PhysRevD.83.085007}
D.E.~Kharzeev and H.-U.~Yee, \emph{Chiral magnetic wave},
  \href{https://doi.org/10.1103/PhysRevD.83.085007}{\emph{Phys. Rev. D}
  {\bfseries 83} (2011) 085007}.

\bibitem{Kogut:1974ag}
J.B.~Kogut and L.~Susskind, \emph{{Hamiltonian Formulation of Wilson's Lattice
  Gauge Theories}}, \href{https://doi.org/10.1103/PhysRevD.11.395}{\emph{Phys.
  Rev.} {\bfseries D11} (1975) 395}.

\bibitem{Susskind:1976jm}
L.~Susskind, \emph{{Lattice Fermions}},
  \href{https://doi.org/10.1103/PhysRevD.16.3031}{\emph{Phys. Rev.} {\bfseries
  D16} (1977) 3031}.

\bibitem{Jordan:1928wi}
P.~Jordan and E.P.~Wigner, \emph{{About the Pauli exclusion principle}},
  \href{https://doi.org/10.1007/BF01331938}{\emph{Z. Phys.} {\bfseries 47}
  (1928) 631}.

\bibitem{Berry:2005yrf}
D.W.~Berry, G.~Ahokas, R.~Cleve and B.C.~Sanders, \emph{{Efficient Quantum
  Algorithms for Simulating Sparse Hamiltonians}},
  \href{https://doi.org/10.1007/s00220-006-0150-x}{\emph{Commun. Math. Phys.}
  {\bfseries 270} (2007) 359}
  [\href{https://arxiv.org/abs/quant-ph/0508139}{{\ttfamily
  quant-ph/0508139}}].

\bibitem{Hatano:2005gh}
N.~Hatano and M.~Suzuki, \emph{{Finding Exponential Product Formulas of Higher
  Orders}}, \href{https://doi.org/10.1007/11526216_2}{\emph{Lect. Notes Phys.}
  {\bfseries 679} (2005) 37}
  [\href{https://arxiv.org/abs/math-ph/0506007}{{\ttfamily math-ph/0506007}}].

\bibitem{Qiskit}
M.S.~Anis et~al., \emph{{Qiskit: An Open-source Framework for Quantum
  Computing}},  {2021}.
\newblock {10.5281/zenodo.2573505}.

\end{thebibliography}\endgroup

\end{document}